\documentclass[draftcls,onecolumn,a4paper]{IEEEtran}

\usepackage{amsmath}
\usepackage{cases}
\usepackage{amsfonts}
\usepackage{amssymb}
\usepackage{boxedminipage}
\usepackage{graphicx}
\usepackage[usenames]{color}
\usepackage{array,color}
\usepackage{colortbl}
\usepackage{extarrows}

\begin{document}
\baselineskip=22pt

\title{The Deterministic Time-Linearity of Service Provided by Fading Channels\footnote{This paper has been published on \textit{IEEE Trans. Wireless Commun.}, May, 2012.}}

\author{Yunquan~Dong, Qing~Wang, Pingyi~Fan,~\IEEEmembership{Senior Member,~IEEE} and~Khaled~Ben~Letaief,~\IEEEmembership{Fellow,~IEEE}
}

\maketitle

\begin{abstract}

In the paper, we study the service process $S(t)$ of an
independent and identically distributed (\textit{i.i.d.})
Nakagami-$m$ fading channel, which is defined as the amount of
service provided, i.e., the integral of the instantaneous channel
capacity over time $t$. By using the Characteristic
Function (CF) approach and the infinitely divisible law, it is proved
that, other than certain generally recognized curve form {or a stochastic process}, the
channel service process $S(t)$ is a deterministic linear
function of time $t$, namely, $S(t)=c_m^\ast\cdot t$ where $c_m^\ast$ is
a constant determined by the fading parameter $m$. Furthermore, we extend it to general \textit{i.i.d.} fading channels and present an explicit form of the constant service rate $c_p^\ast$. The obtained work provides such a new insight on the system design of joint source/channel coding that there exists a
coding scheme such that a receiver can decode with zero error
probability and zero high layer queuing delay, if the transmitter maintains a constant data rate no
more than $c_p^\ast$. Finally, we verify our analysis through Monte
Carlo simulations.

\end{abstract}


\section{Introduction}

In a wireless communication system, signal from a transmitter
usually travels over multiple reflective, diffracted and scattered
paths to a receiver. As a result, the received signal may fluctuate
violently and the received {\em signal to noise ratio} (SNR) varies
randomly over time. This phenomenon is referred to as the multipath
propagation. In particular, when these multipath signals naturally
arrived at the receiver, the fading usually occurs, which is
characterized by Rayleigh, Rician, or the Nakagami-$m$ fading
models. In general, the Nakagami-$m$ fading \cite{MN-nakagami}
includes a wide range of multipath channels via adjusting parameter
$m$. For instance, the Nakagami-$m$ distribution includes the
one-sided Gaussian distribution $(m=1/2)$, which corresponds to the
highest amount of multipath fading scenario. The Rayleigh
distribution is also included by setting $m=1$, which is most
applicable when there is no dominant propagation along the line of
sight (LOS) between a transmitter and the receiver. Moreover, if one
takes $\frac{(K+1)^2}{2K+1}$ as the value of $m$, i.e.,
$m=\frac{(K+1)^2}{2K+1}$, then the Nakagami-$m$ distribution reduces
approximately to the Rician distribution with parameter $K$, which
models the situation when there exists a fixed LOS component in the
received signal\footnote{$K$ is the ratio between the power in the
LOS component and the average non-LOS multipath components
\cite{G-WC}. For $K=0$£¬ it is Rayleigh fading, and for $K=+\infty$
it has no fading.}. In fact, for the modeling of fading channels in
practical communication systems, the Nakagami-$m$ distribution often
gives the best fit to both urban and indoor multipath propagation.

For a wireless link over fading channel, the study of its capacity
has always been of interest. In literature, the {\em Shannon}
({\em ergodic}) {\em capacity} with the channel side information (CSI) at a
receiver for an average power constrain $\overline{P}$ is given in
\cite{MS-CWB}. In this case, the transmission data rate over the channel is
constant regardless of instantaneous SNR at the receiver. The
capacity-achieving code has to be sufficiently long so that a
received codeword is affected by all possible fading states. Besides
the ergodic capacity, the {\em outage capacity} defines the maximal
(fixed) rate achievable in all non-outage states with asymptotically
small error probability \cite{LA-Outage,GS-Throughput}. This capacity usually applies to slowly-varying channels where
instantaneous SNR is constant over a number of transmission slots and
changes to a new value following certain fading distribution.
Moreover, in \cite{GP-CFC}, the authors considered the capacity of
fading channels with side information at both encoder and decoder
with optimal water filling power allocation. In
\cite{BPS-FCI}, it provided an exhaustive review on the
information-theoretic and communication features of fading channels
and derived the capacities of fading channels with and without the
channel side information.

Furthermore, from a viewpoint of cross layer design method, the
authors in \cite{DN-EC} investigated fading channels in terms of the
high layer Qos parameters. According to the large deviation theory
and Legendre transformation, they proposed a link-layer channel
model termed {\em effective capacity} (EC) which specifies the
maximum transmission data rate supported by a fading channel under
certain Qos metrics, such as the maximum delay and the buffer overflow probability. However, their result involves a lot of
inevitable approximation, which limits its application.
Consequently, to facilitate the application of EC theory, the
authors \cite{QDP-EC-R,QDP-EC-N} derived the close form of EC
function for both correlated Rayleigh and Nakagami-$m$ fading channels with the
consideration of Doppler effect and presented the error analysis of
the measurement-based estimation algorithm of the EC function.

In this paper, we first investigate the service process $S(t)$ of the \textit{i.i.d.}
fading channels with the CSI available at the receiver. {Here, the service process $S(t)$ is defined as the} integral of the instantaneous channel capacity over time $t$.
{As the Nakagami-$m$ fading model is widely used and includes several classic fading models by different $m$ parameters, this paper is focused on the characteristics of Nakagami-$m$ fading channels.
For such channels,} the channel magnitude gain varies
randomly following Nakagami-$m$ distribution, the service
process $S(t)$ should also be a stochastic process. However, by
using the infinitely divisible law \cite{Levy} and the CF (Characteristic Function) approach, we prove that the service process
$S(t)$ is itself a deterministic linear function of time $t$, i.e.,
$S(t)=c_m^\ast\cdot t$ where $c_m^\ast$ is a constant determined by the fading parameter $m$.
{Then the result is applied to three special cases of Nakagami-$m$ fading channel, namely, the Rayleigh fading channel, the Rician fading channel and the AWGN channel. Also, the corresponding service rates are derived. Finally,}
we prove that the time
linearity nature of $S(t)$ maintains for all kinds of {\em i.i.d.}
fading channels more than the Rayleigh, Rician and Nakagami fading.
In terms of our analysis, it indicates in theory that there exists a channel coding
scheme such that a constant data rate can be supported by an
\textit{i.i.d.} fading channel with no queuing delay in the
viewpoint of application layer. {Therefore, the \textit{i.i.d.} fading channels have constant and stable transmission ability, just like the AWGN channels.}

In summary, traditionally the ergodic is only considered as one statistic characteristic of fading channels. However, it is proved in this paper that the ergodic capacity is more than that. In fact, {\em i.i.d.} fading channels have the deterministic ability to support a constant traffic rate $c_m^*$, which is equal to the ergodic capacity.
\begin{figure}[!t]
\centering
\includegraphics[width=3.0in]{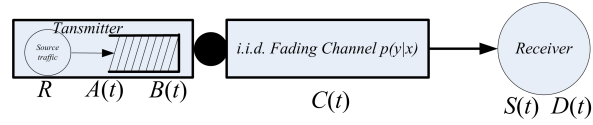}
\caption{The transmission system model} \label{fig:TXm}
\end{figure}
{Particularly, there are two points that should be noted. Firstly, the deterministic time linearity of the service process is related to the fluid traffic model on the large time scales. That is to say, the traffic data is infinitely divisible and we are interested in the channel performance over long periods rather than its performance on the order of symbols. Compared with the time scale interested, the duration when channel gain is constant is very short. Therefore, the channel can be treated as very fast fading channel without loss of generality. However, the duration when channel gain is constant still contains many channel uses so that the $c_p^\ast$ achievable coding scheme can be realized by adapting coding rate according to the channel condition. Secondly, what we are saying is about the deterministic time linearity of $S(t)$ rather than $E[S(t)]$ or $\frac{S(t)}{t}$. Specifically, given an arbitrary time $t_0$, one has $S(t_0)$ is a constant equal to $c_p^\ast \cdot t_0$ instead of a random variable. And as $t_0$ varies, $S(t_0)$ is a deterministic function of $t_0$ instead of a stochastic process. However, the statements that $E[S(t)]$ is a linear function of $t$ and $\frac{S(t)}{t}$ reduces to a constant are straightforward.}

The rest of the paper is organized as follows. In
Section~\ref{sec:model}, we describe the system model, including the
fading channel model. The theoretical analysis is developed in
Section~\ref{sec:analysis} and \ref{sec:general}. More specifically,
we prove the time-linearity of the channel service process $S(t)$
for \textit{i.i.d.} Nakagimi-$m$ fading channels by using the CF
approach and the infinitely divisible law in
Section~\ref{sec:analysis}. Furthermore, we derive the corresponding
linear coefficients i.e., the constant service rate $c^\ast_m$. Then, in Section~\ref{sec:general}, we
show that the deterministic time-linearity nature exists for all
kinds of {\em i.i.d.} fading channels. Numerical results via Monte Carlo
simulations provided in Section~\ref{sec:simulation} to confirm the
time-linearity nature of $S(t)$. This section also provides some related discussions. Finally, we conclude the paper in Section~\ref{sec:conclusions}.

\section{System Model}
\label{sec:model}
\begin{figure}[!t]
\centering
\includegraphics[width=3.0in]{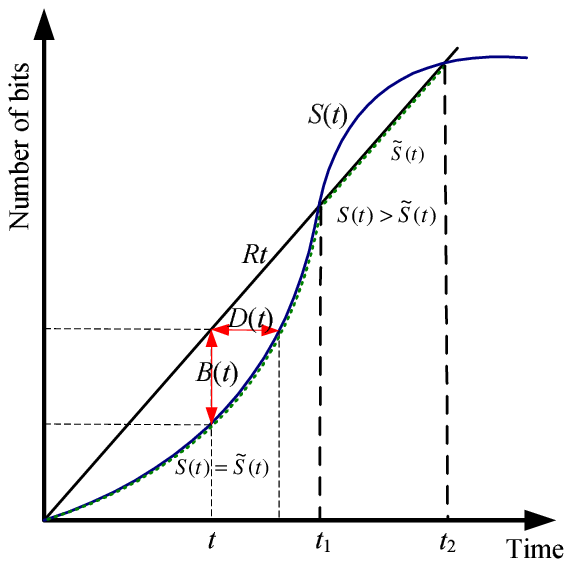}
\caption{Characterization of the source traffic and channel service
process} \label{fig:Service}
\end{figure}

We consider an adaptive transmission model shown in Fig.~\ref{fig:TXm}. {The source
maintains a constant data rate $R$ and the transmitter adapts the transmission rate $C(t)$ according to the channel condition}. The source traffic
stream and channel service ability are matched by a First In First
Out (FIFO) buffer. Let $B(t)$ denote the queuing length in the
buffer in nats at the moment $t$ and $D(t)$ denote the latency that
the data arriving in the system at the moment $t$ will suffer from,
namely, the delay between its arrival time and the moment it has
been served.

Next, we define $S(t)$ as the amount of service provided by the
fading channel until the moment $t$, namely, the solid curve shown
in Fig.~\ref{fig:Service}, and $\tilde{S}(t)$ as the amount of
actual utilized service of the channel until the moment $t$ for the
source traffic, namely, the dotted curve. Intuitively, the
relationship among $B(t)$, $D(t)$, $\tilde{S}(t)$ and $S(t)$ is
depicted in Fig.~\ref{fig:Service}. Note that we have
$\tilde{S}(t)\leq S(t)$ always satisfied. The reason is that, until
any moment $t$, the amount of actually provided service, i.e.,
$\tilde{S}(t)$, should always be jointly upper-bounded by the amount of
source traffic ($Rt$), and the potential amount of service
which the channel can provide, namely, $S(t)$.

By its physical meaning, $S(t)$ is given by
\begin{equation}\label{eq:st}
S(t)=\int_0^t{C(\tau)d\tau},
\end{equation}
where $C(\tau)$ is the instantaneous capacity of the time varying
channel at time $\tau$.

Note that since the instantaneous capacity $C(t)$ is a stochastic
process for $t\geq0$. Therefore, $S(t)$ is also a stochastic process which is
equivalent to the stochastic integral in the sense of zero
mean-square error, based on the stochastic calculus theory. That is,
for any time $t\geq0$, the following relation holds,
\begin{equation}\label{eq:mean_squ_equ}
E[(S(t)-\int_0^t{C(\tau)d\tau} )^2]=0.
\end{equation}

However, one amazing thing is that it will be proved that $S(t)$ has
time-linearity nature which means its profile is a line, other than
certain generally recognized curve shown in Fig.~\ref{fig:Service}.

In particular, for the channel service process $S(t)$ associated
with the instantaneous channel capacity $C(t)$ mentioned above, we
consider a continuous time Nakagami-$m$ fading channel with the
stationary and ergodic time varying channel gain $g(t)$ and additive
white Gaussian noise (AWGN) $n(t)$. For any moment $t_0$, the
channel gain $g(t_0)$ is Nakagami-$m$ distributed and the
corresponding power gain is denoted by $\gamma(t_0)=g^2(t_0)$. For $m\geq \frac12$, their {\em probability density functions} ({\em p.d.f}) are given by
\begin{subequations}
\begin{equation}\label{eq:nakagami}
p^{m}_{g}(g)=2{(\frac{m}{P_r})}^m \frac{g^{2m-1}}{\Gamma(m)}
e^{\frac{-mg^2}{P_r}},
\end{equation}
\begin{equation}\label{eq:nakagami2}
p^{m}_{\gamma}(\gamma)=(\frac{m}{P_r})^m \frac{\gamma^{m-1}}{\Gamma(m)} e^{\frac{-m\gamma}{P_r}},
\end{equation}
\end{subequations}
where $g(t)$ is independent of both the channel input process and
any other $g(t')$ for $t\neq t'$, and so is $\gamma(t)$. Here, $P_r$
is the average received signal power {of a receiver unit distance away} and $\Gamma(\cdot)$ is the
Gamma function. Let $P_t$ denote the average transmit power of the
signal, and $N_0$ denote the noise power spectral density. $W$ is
the limited bandwidth of the received signal. Note that, without
loss of generality, let $d$ and $\alpha$ denote the distance between
the transmitter and receiver and the path loss exponent, respectively. Then, the
instantaneous received SNR is given by
\begin{equation}\label{eq:snr}
{\rm SNR}=\frac{\gamma(t)P_td^{-\alpha}}{WN_0},
\end{equation}
and the corresponding instantaneous capacity of the fading channel
in $nats$ is given by
\begin{equation}\label{eq:insC}
C(t)=W\ln(1+\frac{\gamma(t)P_td^{-\alpha}}{WN_0}).
\end{equation}

In particular, for $m=1$, the Nakagami-$m$ distribution is reduced to the
Rayleigh distribution, which is given by
\begin{subequations}
\begin{equation}\label{eq:rayle}
p^{ra}_{g}(g)=\frac{2g}{P_r}e^{\frac{-g^2}{P_r}},
\end{equation}
\begin{equation}\label{eq:expo}
p^{ra}_{\gamma}(\gamma)=\frac{1}{P_r}e^{\frac{-\gamma}{P_r}}.
\end{equation}
\end{subequations}

The Rician distribution can also be approximated by the Nakagami-$m$
distribution by setting $m=\frac{(K+1)^2}{2K+1}$. The Rician
distribution in terms of parameter $K$ is given by{\small
\begin{eqnarray}\label{rician}
p^{ri}_{g}&=&\frac{2g(K+1)}{P_r} \exp [-K-\frac{(K+1)g^2}{P_r}] \nonumber \\
&&I_0 (2g\sqrt{ \frac{K(K+1)}{P_r} }),~ ~ ~ g\geq0,
\end{eqnarray}
}
where $K=\frac{s^2}{\sigma^2}$ is the ratio between the power in the
LOS component and the average non-LOS multipath components.

Upon all of the preparation above, we will begin to analyze the
service process $S(t)$ and study its property next.

\section{Time-Linearity of $S(t)$ for Nakagami Fading Channels}\label{sec:analysis}

In this paper, we are interested in the statistic characteristics
of $S(t)$ and study some properties of such a process by exploiting
its Characteristic Function (CF). Specifically, for an \textit{i.i.d.}
Nakagami-$m$ fading channel, we give the following theorem which is
one of the main contributions of the paper.

\textbf{\textit{Theorem} 1}: \textit{For an i.i.d. Nakagami-$m$ fading
channel, the amount of service provided by the channel during time
$t$, namely, $S(t)$, is a deterministic linear function of time $t$,
i.e., $S(t)=c_m^\ast\cdot t$, where $c_m^\ast$ is a constant and given by}
\begin{equation}\label{eq:st}
c_m^\ast=\frac{W}{\Gamma(m)} \int_0^\infty \ln
(\frac{P_rP_t}{mWN_0d^\alpha} y+1) y^{m-1} e^{-y}dy,
\end{equation}
\textit{where $P_r$ is the average received signal power of the
signal. $N_0$ is the noise power spectral
density, $W$ is the received signal bandwidth, $P_t$ is the average
transmit power and $d$ is the distance between the transmitter and
receiver. The function $\Gamma(m)=\int_0^\infty t^{m-1}e^{-t}dt$ is
the Gamma function.}

First, note that, $c_m^\ast$ is not the instantaneous capacity of
the channel, namely, $C(t)$, which means that we have $c_m^\ast\neq
\textrm{d}S(t)/\textrm{d}t$. However, there does not exist any contradiction.
Actually, $S(t)$ is continuous and non-differentiable  in $[0,t]$,
just like the Brown motion process, which is an integration of a
Gaussian process. It is easy to understand since the signal obeying
Nakagami-$m$ distribution can be viewed as the square root of the
summation of $2m$ squared independent Gaussian random variables.
Hence, it has  the similar property as Brown motion naturally.

To prove {\em Theorem} 1, we first show that $S(t)$ is a Levy
Process. Then, its CF is derived based on the infinitely divisible
law for the Levy process. With the CF, the theorem can be proved
then.

\subsection{Review of the Levy Process and Infinitely
Divisibility}\label{subsec:levy and divisibility}

We first review the definition of \textit{Levy process and
infinitely divisibility} as below.

\textbf{\textit{Definition} 1} \cite{Levy}:
$X=\{X(t)\}_{t\geq0}$ is said to be a \textit{Levy process} if
\begin{enumerate}
  \item $X$ has independent increments.
  \item $X(0)$=0 is satisfied almost surely.
  \item $X$ is stochastically continuous, i.e., for $s\geq0$, $X(t+s)-X(s)\xrightarrow{P}0$ as $t\rightarrow0$.
  \item $X$ is time homogeneous, i.e., for $t\geq0$, $\mathcal{L}(X(t+s)-X(s))$ does not depend on $s\geq0$.
  \item $X$ is Right Continuous with Left Limits (RCLL) almost surely.
\end{enumerate}

\textbf{\textit{Definition} 2} \cite{Id}: a probability
distribution $F$ of a random variable on the real line is
\textit{infinitely divisible}, if, for every positive integer $n$,
there exists $n$ \textit{i.i.d.} random variables $X_1, ..., X_n$
whose sum is equivalent to $X$ in distribution. Note that these $n$
random variables do not have to obey the same probability
distribution as $X$.

According to the definition of \textit{Levy process and infinitely
divisibility}, the condition of the infinitely divisibility for a
Levy process is given by the following proposition.

\textbf{\textit{Proposition} 1}: \textit{For a random vector $Y$ in
$\mathbb{R}^d$, the following three statements are equivalent} (Theorem 1.3 \cite{Levy}).
\begin{enumerate}
\item $Y$ is infinitely divisible.
\item $Y_{n,1}+\cdots+Y_{n,r_{n}}\xrightarrow{d}Y$ for some \textit{i.i.d.} array $(Y_{n,j})_{n\geq1,r_n\geq j\geq1}$, where $r_n\rightarrow\infty$.
\item $Y\xlongequal{d} X_1$ for some Levy process $X$ in
$\mathbb{R}^d$.
\end{enumerate}

Next, we shall show that the service process $\{S(t)\}_{t\geq0}$
satisfies the five conditions given in \textit{Definition} 1 and it
is a Levy process. First, for the channel magnitude gain $g(t)$
which is \textit{i.i.d.} in our discussion, the service process
$S(t)$ has independent increments. Second, the condition $S(0)=0$ is
easily satisfied. Third, it is clear that $S(t)$ is stochastically
continuous due to the condition
$X(t+s)-X(s)=\int_s^{s+t}C(\tau)d\tau\xrightarrow{P}0$ as
$t\rightarrow0$. Besides, $S(t)$ is also time homogeneous since its
components $C(\tau)$ is \textit{i.i.d.} over $\tau$. Finally, it is
easy to see that $S(t)$ is Right Continuous with Left Limits (RCLL),
since there is no leap in $S(t)$ and for any $t\ge0$, $S(t)$ exists.

Thus, according to \textit{Proposition} 1, the distribution of
$S(t)$ is infinite divisible and, for any $n>0$, $S(t)$ can be
decomposed into the sum of $n$ \textit{i.i.d.} random variables.

Now, for sufficiently large $N$, we define the time resolution
$\Delta\tau=\frac{t}{N}$ and obtain the samples $\Delta
s_n=C(t_n)\Delta t$ where $t_n=n\Delta\tau, n=0,1,\cdots,N$. Then,
we get
\begin{equation}\label{eq:st_divis}
S(t)=\lim_{N\rightarrow\infty}\sum_{n=0}^{N}C(t_n)\Delta
t=\lim_{N\rightarrow\infty}\sum_{n=0}^{N}\Delta s_n,
\end{equation}
where $\Delta s_n$ are \textit{i.i.d.} random variables by the same probability distribution, which will be
given in Section~\ref{subsec:CDF_sn}.

\subsection{Derivation of the CDF of $C(t)$ and $\Delta
s_n$}\label{subsec:CDF_sn}

To develop the CF of $C(t)$ and $\Delta s_n$, we need to derive
their Cumulative Distribution Function (CDF) first. More specifically, for the CDF of $C(t)$, it is
given by
\begin{equation}\label{eq:Fct}
\begin{split}
F_{C}(c)&=P\{C(t)<c\}\\
&=P\{\gamma(t)<\frac{WN_0d^\alpha}{P_t}(e^{\frac{c}{W}}-1)\}\\
&=\frac{1}{\Gamma(m)}(\frac{m}{P_r})^m
\int_0^{\frac{WN_0d^\alpha}{P_t}(e^{\frac{c}{W}}-1)}
\gamma^{m-1}e^{-\frac{m}{P_r}\gamma}d\gamma,
\end{split}
\end{equation}
and, for the CDF of $\Delta s_n$, it is given by
\begin{equation}\label{eq:Fsn}
\begin{split}
F_{\Delta s_n}(\Delta s)&=P\{\Delta s_n<\Delta s\}\\
&=F_{C}(\frac{\Delta s}{\Delta\tau})\\
&=\frac{1}{\Gamma(m)}(\frac{m}{P_r})^m
\int_0^{\frac{WN_0d^\alpha}{P_t}(e^{\frac{\Delta
s}{W\Delta\tau}}-1)} \gamma^{m-1}e^{-\frac{m}{P_r}\gamma}d\gamma.
\end{split}
\end{equation}

According to the CDF expression of $\Delta s_n$ given in
Eqn.(\ref{eq:Fsn}), we introduce the following proposition.

\textbf{\textit{Proposition} 2}: \textit{For an arbitrary $\varepsilon>0$,
the equation below
\begin{equation}\label{eq:propos2}
\lim_{N\rightarrow\infty}P\{\Delta s_n>\varepsilon \} = 0
\end{equation}
is always satisfied.}

\textit{Proof}: According to the CDF of $\Delta s_n$ in
Eqn.(\ref{eq:Fsn}), we get
\begin{equation}\label{eq:Fsn_proof}
\begin{split}
&\lim_{N\rightarrow\infty}P\{\Delta s_n>\varepsilon\}\\
=&\lim_{N\rightarrow\infty}\{1-F_{\Delta s_n}(\varepsilon)\}\\
=&\lim_{\Delta\tau\rightarrow0}\{ 1- \frac{1}{\Gamma(m)}(\frac{m}{P_r})^m \int_0^{\frac{WN_0d^\alpha}{P_t}(e^{\frac{\varepsilon}{W\Delta\tau}}-1)} \gamma^{m-1}e^{-\frac{m}{P_r}\gamma}d\gamma\}\\
=&0.
\end{split}
\end{equation}

Thus, no leap exists at any time $t$ for $S(t)$, which indicates
that $S(t)$ is continuous in mean-square right continuous with left limits. $\square$

Until now, with the CDF of $\Delta s_n$, we can derive the CF of
$S(t)$ in Section~\ref{subsec:CF_sn}.

\subsection{Derivation of the CF of $S(t)$}\label{subsec:CF_sn}

First,let us calculate the CF of $\Delta s_n$ as follows,{\small
\begin{equation}\label{eq:CF_sn}
\begin{split}
&\varphi_{\Delta s_n}(\lambda)=E[e^{i\lambda \Delta s}]\\
=&\int_0^\infty e^{i\lambda \Delta s} dF_{\Delta s_n}(\Delta s) \\
\stackrel{(a)}{=}&\frac{1}{\Gamma(m)} (\frac{mWN_0d^\alpha}{P_rP_t})^m \int_1^\infty x^{i\lambda W\Delta\tau} (x-1)^{m-1} e^{ -\frac{mWN_0d^\alpha}{P_rP_t}(x-1) } dx\\
\stackrel{(b)}{=}&\frac{1}{\Gamma(m)}
(\frac{mWN_0d^\alpha}{P_rP_t})^m \int_0^\infty (y+1)^{i\lambda
W\Delta\tau} y^{m-1} e^{ -\frac{mWN_0d^\alpha}{P_rP_t}y } dy,
\end{split}
\end{equation}}
where some variable substitutions are used, e.g. $x=e^{\frac{\Delta
s}{W\Delta\tau}}$ in (a) and $y=x-1$ in (b), respectively. Note that
$i=\sqrt{-1}$ is the imaginary unit and $\Gamma(m)=\int_0^\infty
t^{m-1}e^{-t}dt$ is the Gamma Function.

Next, as $N\rightarrow\infty$, we have the following result.{\small
\begin{equation}\label{eq:yili}
\begin{split}
&\lim_{N\rightarrow\infty}\frac{1}{\Gamma(m)} (\frac{mWN_0d^\alpha}{P_rP_t})^m \int_0^\infty (y+1)^{i\lambda W\Delta\tau} y^{m-1} e^{ -\frac{mWN_0d^\alpha}{P_rP_t}y } dy\\
=&\lim_{\Delta\tau\rightarrow0}\frac{1}{\Gamma(m)} (\frac{mWN_0d^\alpha}{P_rP_t})^m \int_0^\infty (y+1)^{i\lambda W\Delta\tau} y^{m-1} e^{ -\frac{mWN_0d^\alpha}{P_rP_t}y } dy\\
=&\frac{1}{\Gamma(m)} (\frac{mWN_0d^\alpha}{P_rP_t})^m \int_0^\infty y^{m-1} e^{ -\frac{mWN_0d^\alpha}{P_rP_t}y } dy\\
\stackrel{(a)}{=}&\frac{1}{\Gamma(m)} (\frac{mWN_0d^\alpha}{P_rP_t})^m(\frac{P_rP_t}{mWN_0d^\alpha})^m \Gamma(m)\\
=&1,
\end{split}
\end{equation}
}
where the variable substitution $z=\frac{mWN_0d^\alpha}{P_rP_t}y$ is
used in (a).

With the CF of $\Delta s_n$, to derive the CF of $S(t)$, we first give the following lemma.

\textbf{\textit{Lemma} 1}: \textit{The following item}
\begin{equation*}
\frac{1}{\Gamma(m)} (\frac{mWN_0d^\alpha}{P_rP_t})^m \int_0^\infty
(y+1)^{i\lambda w\Delta\tau} y^{m-1} e^{
-\frac{mWN_0d^\alpha}{P_rP_t}y } dy-1 \nonumber
\end{equation*}
\textit{is an infinitesimal of the same order with $\Delta\tau$,
where $\Delta\tau=\frac{t}{N}$.}

\textit{Proof}: To prove one is an infinitesimal of the same order
with the other, we need to compute {\small
\begin{equation*}\label{eq:order}
\lim_{\Delta\tau\rightarrow0} \frac{\frac{1}{\Gamma(m)}
(\frac{mWN_0d^\alpha}{P_rP_t})^m \int_0^\infty (y+1)^{i\lambda
W\Delta\tau} y^{m-1} e^{ -\frac{mWN_0d^\alpha}{P_rP_t}y }
dy-1}{\Delta\tau} \nonumber
\end{equation*}
\begin{eqnarray}
&&=\lim_{\Delta\tau\rightarrow0} \frac{1}{\Gamma(m)} (\frac{mWN_0d^\alpha}{P_rP_t})^m \times \nonumber\\
&&\int_0^\infty i\lambda W (y+1)^{i\lambda W\Delta\tau} \ln(y+1) y^{m-1} e^{ -\frac{mWN_0d^\alpha}{P_rP_t}y } dy \nonumber \\
&\stackrel{(a)}{=}&\frac{1}{\Gamma(m)} (\frac{mWN_0d^\alpha}{P_rP_t})^m \cdot i\lambda W \int_0^\infty \ln(y+1) y^{m-1} e^{ -\frac{mWN_0d^\alpha}{P_rP_t}y } dy \nonumber \\
&\stackrel{(b)}{=}&\frac{1}{\Gamma(m)} \cdot i\lambda W \int_0^\infty \ln(\frac{P_rP_t}{mWN_0d^\alpha}z+1) z^{m-1} e^{ -z } dz \nonumber \\
&\stackrel{(c)}{=}&i\lambda W c_0 ,
\end{eqnarray}
}
where, in (a) we apply the L'H\^{o}pital's rule and in (b), we use
variable substitutions $z=\frac{mWN_0d^\alpha}{P_rP_t}y$ and in (c),
\begin{equation}\label{eq:c0}
c_0=\frac{1}{\Gamma(m)} \int_0^\infty
\ln(\frac{P_rP_t}{mWN_0d^\alpha}z+1) z^{m-1} e^{ -z } dz.
\end{equation}

If $c_0$ is finite, then the limit in (b) should also be finite and
the proof is completed. This can be assured by {\em Lemma} 2.
\hspace{10cm} $\square$

\textbf{\textit{Lemma} 2}: \textit{$c_0$ is finite and its lower and upper
bound are given by}
\begin{equation}\label{eq:c0_bound}
    a\le c_0 \le \frac{P_rP_t}{WN_0d^\alpha},
\end{equation}
where $\Gamma(s,x)=\int_x^\infty t^{s-1}e^{-t}dt$ is the incomplete Gamma Function and $a=\frac{\Gamma(m, 1)}{\Gamma(m)} \ln(\frac{P_rP_t}{mWN_0d^\alpha} +1)$.

\textit{Proof}: It is easy to see that $c_0$ is a finite number if it is finitely lower and upper bounded. Firstly, for its lower bound we have

\begin{subequations}\label{eq:finitec0}
\begin{equation}\label{eq:bigger}
\begin{split}
c_0=&\frac{1}{\Gamma(m)} \int_0^\infty \ln(\frac{P_rP_t}{mWN_0d^\alpha}z+1) z^{m-1} e^{ -z } dz \\
>&\frac{1}{\Gamma(m)} \int_1^\infty \ln(\frac{P_rP_t}{mWN_0d^\alpha} z+1) z^{m-1} e^{ -z } dz\\
\stackrel{(a)}{\geq}&\frac{1}{\Gamma(m)} \int_0^\infty \ln(\frac{P_rP_t}{mWN_0d^\alpha}\cdot 1 +1) z^{m-1} e^{ -z } dz\\
=&\frac{1}{\Gamma(m)} \Gamma(m, 1)\ln(\frac{P_rP_t}{mWN_0d^\alpha} +1)\\
\stackrel{d}{=}&a,
\end{split}
\end{equation}
where (a) comes from the fact that $z\geq1$.

Similarly, for its upper bound we have
\begin{equation}\label{eq:smaller}
\begin{split}
c_0=&\frac{1}{\Gamma(m)} \int_0^\infty \ln(\frac{P_rP_t}{mWN_0d^\alpha}z+1) z^{m-1} e^{ -z } dz \\
\stackrel{(a)}{\leq}&\frac{1}{\Gamma(m)} \int_0^\infty \frac{P_rP_t}{mWN_0d^\alpha}z\cdot z^{m-1} e^{ -z } dz\\
=&\frac{P_rP_t}{mWN_0d^\alpha} \frac{\Gamma(m+1)}{\Gamma(m)} \\
\stackrel{(b)}{=}&\frac{P_rP_t}{WN_0d^\alpha},
\end{split}
\end{equation}
\end{subequations}
where we have (a) by inequality $\ln(\frac{P_rP_t}{mWN_0d^\alpha}z+1)\leq
\frac{P_rP_t}{mWN_0d^\alpha}z$ and (b) follows
$\Gamma(m+1)=m\Gamma(m)$. Based on \textit{Lemma} 1 and \textit{Lemma} 2, it is assured that
$(\frac{1}{\Gamma(m)} (\frac{mWN_0d^\alpha}{P_rP_t})^m \int_0^\infty
(y+1)^{i\lambda w\Delta\tau} y^{m-1}e^{
-\frac{mWN_0d^\alpha}{P_rP_t}y } dy-1)$ and $\Delta\tau$ are of the
same order. $\square$

Finally, by using the properties given by (\ref{eq:yili}) and
(\ref{eq:order}) in {\em Lemma} 1, the CF of $S(t)$ can be derived
as follows.
\begin{equation}\label{eq:CFst}
\begin{split}
&\varphi_{s(t)}(\lambda)=E[e^{i\lambda s(t)}]\\
=&\lim_{N\rightarrow\infty} [\varphi_{\Delta s_n}(\lambda)]^N\\
=&\lim_{\Delta\tau\rightarrow0} [\varphi_{\Delta s_n}(\lambda)]^{\frac{t}{\Delta\tau}}\\
=&\lim_{\Delta\tau\rightarrow0}[ \frac{1}{\Gamma(m)} (\frac{mWN_0d^\alpha}{P_rP_t})^m \cdot\\
&~~~~~~~\int_0^\infty (y+1)^{i\lambda W\Delta\tau} y^{m-1} e^{ -\frac{mWN_0d^\alpha}{P_rP_t}y } dy ]^{\frac{t}{\Delta\tau}}\\
\stackrel{(a)}{=}&\lim_{\Delta\tau\rightarrow0}[ 1+i\lambda W c_0\Delta\tau +o(\Delta\tau)]^{\frac{t}{\Delta\tau}}\\
\stackrel{(b)}{=}&e^{i\lambda W c_0t}
\end{split}
\end{equation}
where in (a), we use the property given by (\ref{eq:order}) and (b)
follows the known result
$\lim_{x\rightarrow0}(1+kx)^{\frac{1}{x}}=e^k$.

Besides, the relationship between the moments of a random
variable $X$ and its CF $\varphi_X(\lambda)$ is given by
\begin{subequations}\label{eq:rcrela}
\begin{equation}
E(X)=\frac{1}{i}\varphi'_X(\lambda)|_{\lambda=0},
\end{equation}
\begin{equation}\label{eq:rcrela_square}
E(X^2)=\frac{1}{i^2}\varphi{''}_X(\lambda)|_{\lambda=0}.
\end{equation}
\end{subequations}
Then we get the following numerical characteristics of $S(t)$ directly,
\begin{subequations}\label{eq:meanst_varst}
\begin{equation}\label{eq:meanst_a}
E(S(t))=\frac{1}{i}\varphi'_{S(t)}(\lambda)|_{\lambda=0}=Wc_0t,
\end{equation}
\begin{equation}\label{eq:varst_b}
E(S^2(t))=\frac{1}{i^2}\varphi{''}_{S(t)}(\lambda)|_{\lambda=0}= [Wc_0t]^2.
\end{equation}
\end{subequations}

By using the results given by (\ref{eq:meanst_a}) and
(\ref{eq:varst_b}), the variance of $S(t)$ is
\begin{equation}\label{eq:varst}
D(S(t))=E(S^2(t))-E^2(S(t))=0.
\end{equation}

It is clear that for any given $t$, $S(t)$ is a random variable
with zero variance, namely,

\begin{equation}\label{eq:result}
    S(t)=E[S(t)]=Wc_0t.
\end{equation}

This means that it is a deterministic linear
function of $t$ in accord with the expression in
(\ref{eq:result}), where $c_0$ is given by (\ref{eq:c0}) and the
linear coefficient $c_m^\ast$ is given by (\ref{eq:st}) in {\em
Theorem} 1. Up to now, we complete the proof of {\em Theorem} 1.

Next, based on {\em Theorem} 1, we investigate three special cases of Nakagami-$m$ fading channel, namely $m=1$, $m=\frac{(K+1)^2}{2K+1}$ and $m=\infty$, which corresponds to the Rayleigh fading channel, the Rician fading channel and the channel with no fading, respectively.

Firstly, for the \textit{i.i.d.} Rayleigh fading channel, we have the following corollary.

\textbf{\textit{Corollary} 1}: \textit{For an i.i.d. Rayleigh fading channel,
the service process $S(t)$ is a deterministic linear function of time $t$ given by $S(t)=c_{ra}^\ast\cdot t$, where $c_{ra}^\ast$ is a constant and given by}
\begin{equation}\label{eq:st_rayleigh}
c_{ra}^\ast=W e^{\frac{WN_0d^\alpha}{P_rP_t}}
\textmd{Ei}(\frac{WN_0d^\alpha}{P_rP_t}),
\end{equation}
\textit{where $P_r$ is the average received signal power, $N_0$ is
the noise power spectral density, $W$ is the received signal
bandwidth, $P_t$ is the average transmit power and $d$ is the
distance between the transmitter and receiver. The function}

\begin{equation*}
\textmd{Ei}(x)=\int^{+\infty}_x {\frac{e^{-x}}{x}dx}=\eta+\ln
x-x+\frac{1}{2}\frac{x^2}{2!}-\frac{1}{3}\frac{x^3}{3!}+\cdots
\nonumber
\end{equation*}
 \textit{is the exponential integration where the {\em Euler's} constant
is $\eta=\int_{+\infty}^0 {\frac{e^{-x}}{x}dx}=0.577215655$.}

\textit{Proof}: It is known that Rayleigh distribution is a special
case of Nakagami distribution for $m=1$. Then, according to
(\ref{eq:st}), we have
\begin{equation}\label{st_rayleigh}
\begin{split}
    c_{ra}^{\ast}&=c_m^{\ast}|_{m=1} \\
    &=\{\frac{W}{\Gamma(m)} \int_0^\infty \ln (\frac{P_rP_t}{mWN_0d^\alpha} y+1) y^{m-1} e^{-y}dy\}|_{m=1} \\
    &\stackrel{(a)}{=}W\cdot\int_0^\infty \ln (\frac{P_rP_t}{WN_0d^\alpha} y+1) e^{-y}dy \\
    &\stackrel{(b)}{=}W\frac{P_rP_t}{WN_0d^\alpha} \int_0^\infty e^{-y} \frac{1}{\frac{P_rP_t}{WN_0d^\alpha}y+1}dy\\
    &\stackrel{(c)}{=}We^{\frac{WN_0d^\alpha}{P_rP_t}} \int_{\frac{WN_0d^\alpha}{P_rP_t}}^\infty \frac{e^{-z}}{z}dz\\
    &=-We^{\frac{WN_0d^\alpha}{P_rP_t}} \textrm{Ei}(\frac{WN_0d^\alpha}{P_rP_t}),
\end{split}
\end{equation}
where (a) follows $\Gamma(1)=1$. We apply the integration by
parts in (b) and variable substitution
$z=y+\frac{P_rP_t}{WN_0d^\alpha}$ in (c). \hspace{12cm} $\square$

Secondly, for the \textit{i.i.d.} Rician fading case, we have the following corollary.

\textbf{\textit{Corollary} 2}: \textit{For an i.i.d. Rician fading channel,
the service process $S(t)$ is a deterministic linear function of time $t$ given by $S(t)=c_{ri}^\ast\cdot t$, where $c_{ri}^\ast$ is a constant and given by}
\begin{equation}\label{eq:st_rician}
c_{ri}^\ast=\frac{W}{\Gamma(\frac{(K+1)^2}{2K+1})} \int_0^\infty
\ln(\frac{P_rP_t}{ \frac{(K+1)^2}{2K+1} WN_0d^\alpha}y+1 )
y^{\frac{K^2}{2K+1}} e^{-y} dy ,
\end{equation}
\textit{where Rician parameter $K=\frac{s^2}{\sigma^2}$ is the ratio
between the power in the LOS component and the average non-LOS
multipath components, $N_0$ is the noise power spectral density, $W$
is the received signal bandwidth, $P_t$ is the average transmit
power and $d$ is the distance between the transmitter and receiver.
}

For this corollary, we simply substitute $m=\frac{(K+1)^2}{2K+1}$ in (\ref{eq:st}) and this complets the proof. Note that we did not get an explicit closed form expression here. Some upper and lower bounds may be needed. We shall discuss it in the future.

Furthermore, for the case when there is no fading, we have the following corollary.

\textbf{\textit{Corollary} 3}: \textit{For an i.i.d. Nakagami-$m$ fading channel with negligible fading, i.e., $m\rightarrow\infty$, the service process $S(t)$ is a deterministic linear function of time $t$ given by $S(t)=c_{\infty}^\ast\cdot t$, where $c_{\infty}^\ast$ is a constant and given by}
\begin{equation}\label{eq:st_infty}
c_{\infty}^\ast=W\ln(1+\frac{P_rP_t}{WN_0d^\alpha}),
\end{equation}
\textit{where $P_r$ is the average received signal power, $N_0$ is
the noise power spectral density, $P_t$ is the average transmit
power and $d$ is the distance between the transmitter and receiver.
}

\textit{Proof}: Let $m\rightarrow\infty$ in (\ref{eq:st}), we have
\begin{equation}\label{eq:st_infty_dr}
    \begin{split}
    c_{\infty}^{\ast}&=c_m^{\ast}|_{m\rightarrow\infty} \\
    &=\lim_{m\rightarrow\infty} \frac{W}{\Gamma(m)} \int_0^\infty \ln (\frac{P_rP_t}{mWN_0d^\alpha} y+1) y^{m-1} e^{-y}dy \\
    &\stackrel{(a)}{\leq}\lim_{m\rightarrow\infty} W \ln( \frac{P_rP_t}{mWN_0d^\alpha} \frac{1}{\Gamma(m)} \int_0^\infty y\cdot y^{m-1} e^{-y}dy  +1)\\
    &=\lim_{m\rightarrow\infty} W \ln( \frac{P_rP_t}{WN_0d^\alpha} \frac{1}{m\Gamma(m)} \Gamma(m+1) +1)\\
    &\stackrel{(b)}{=}W\ln(1+\frac{P_rP_t}{WN_0d^\alpha}),
    \end{split}
\end{equation}
where (a) follows the Jensen' inequality and the unction $\ln(\frac{P_rP_t}{mWN_0d^\alpha} y +1)$ is an concave function of $y$ and (b) follows
$\Gamma(m+1)=m\Gamma(m)$.

However, we can see from (\ref{eq:nakagami2}) that
\begin{equation}\label{nakagami_limit}
\begin{split}
    &\lim_{m\rightarrow\infty} p^{(m)}_{\gamma}=\delta(\gamma-P_r).
\end{split}
\end{equation}

It is easy to understand because  no fading exists when
$m\rightarrow\infty$ and the channel power gain is a constant
$\sqrt{P_r}$ between the transmitter and the receiver. Hence, in (a)
of (\ref{eq:st_infty_dr}), the equality holds. \hspace{1cm} $\square$

It is worthy to be noted that this result can be predicated intuitively in AWGN channel, which further comfired our developed theory in {\em Theorem} 1.

\section{Time-linearity of General \textit{i.i.d.} Fading Channels}\label{sec:general}

In this section, we will show that the deterministic time-linearity is true
for arbitrary \textit{i.i.d.} fading channels.

Assume an \textit{i.i.d.} fading channel with its power gain
$\gamma(t)$ following distribution $p_\gamma(\gamma)$. Let $C(t)$
and $S(t)$ be the instantaneous channel capacity and the service
process, we have the following theorem.

\textbf{\textit{Theorem} 2}: \textit{For an arbitrary i.i.d. fading channel with its {\em p.d.f.} $p_\gamma(\gamma)$,
the service process $S(t)$ is a deterministic linear function of time $t$ given by $S(t)=c_p^\ast\cdot t$, where $c_p^\ast$ is a constant and given by}
\begin{equation}\label{eq:st_p}
c_p^\ast=W \int_0^\infty \ln (\frac{P_t}{WN_0d^\alpha} y+1)
p_\gamma(y) dy,
\end{equation}
\textit{where $N_0$ is the noise power spectral density, $W$ is the
received signal bandwidth, $P_t$ is the average transmit power and
$d$ is the distance between the transmitter and receiver. }

The proof of {\em Theorem} 2 is similar to that of {\em Theorem} 1 and its sketch is provided below.

Firstly, we get the CDF of $C(t)$ and $\Delta s_n$ by
\begin{subequations}\label{general_cs}
\begin{equation}\label{general_c}
    F_{C}(c)=\int_0^{\frac{WN_0d^\alpha}{P_t}( e^{\frac{c}{W}}-1 )} p_\gamma(\gamma)d\gamma,
\end{equation}
\begin{equation}\label{general_s}
    F_{\Delta s_n}(\Delta s)=\int_0^{\frac{WN_0d^\alpha}{P_t}( e^{\frac{\Delta s}{W\Delta \tau}}-1 )} p_\gamma(\gamma)d\gamma.
\end{equation}
\end{subequations}

Then we derive the CF of $\Delta s_n$.

\begin{equation}\label{CF_snp}
\begin{split}
\varphi_{\Delta s_n}(\lambda)&=E(e^{i\lambda\Delta s_n})\\
&\stackrel{(a)}{=}\int_0^\infty e^{i\lambda\Delta s} d F_{\Delta s_n}(\Delta s)\\
&\stackrel{(b)}{=}\frac{WN_0d^\alpha}{P_t}\int_1^\infty x^{i\lambda
W\Delta\tau}p_\gamma( \frac{WN_0d^\alpha}{P_t}(x-1) )dx,
\end{split}
\end{equation}
where in (b) we apply the variable substitution $x=\frac{\Delta s}{W\Delta\tau}$.

From the step (a) in (\ref{CF_snp}), it is easy to see that
$\lim_{\Delta\tau\rightarrow0}\varphi_{\Delta s_n}(\lambda)=1$. With
the similar computational procedure as (\ref{eq:order}), we have
\begin{equation}\label{order_p}
\lim_{\Delta\tau\rightarrow0}\frac{\varphi_{\Delta
s_n}(\lambda)-1}{\Delta\tau}=i\lambda W\int_0^\infty
\ln(\frac{P_t}{WN_0d^\alpha}y+1) p_\gamma(y)dy.
\end{equation}

Finally, the CF of $S(t)$ can be derived as
\begin{equation}\label{CF_P}
\begin{split}
    &\varphi_{S(t)}(\lambda)=\lim_{N\rightarrow\infty} [\varphi_{\Delta s_n}(\lambda)]^N\\
    =&\lim_{\Delta\tau\rightarrow0} [\varphi_{\Delta s_n}(\lambda)]^{\frac{t}{\Delta\tau}}\\
    =&\lim_{\Delta\tau\rightarrow0} [1+ i\lambda W\Delta\tau \int_0^\infty \ln(\frac{P_t}{WN_0d^\alpha}y+1) p_\gamma(y)dy +o(\Delta\tau)]^{\frac{t}{\Delta\tau}}\\
    =&\exp[i\lambda Wt \int_0^\infty \ln(\frac{P_t}{WN_0d^\alpha}y+1) p_\gamma(y)dy].
\end{split}
\end{equation}

According to (\ref{eq:rcrela}), we get $E(S(t))=c_p^\ast \cdot t$ and $D(S(t))=0$, where $c_p^\ast$ is given by (\ref{eq:st_p}). This means that $S(t)=c_p^\ast\cdot t$ and completes the proof of {\em Theorem} 2.

Up to now, we have proved that the deterministic time-linearity
nature exists for all kinds of {\em i.i.d.} fading channels and also
derived the linear coefficients i.e., the constant service rate
$c^\ast_p$, for all kinds of fading channels.

\section{Numerical Results and Discussions}\label{sec:simulation}

To demonstrate the time-linearity of the channel service process
$S(t)$, we consider a point to point communication system over an
\textit{i.i.d.} fading channel, as shown in Fig.~\ref{fig:TXm},
where the average LOS received power $P_r$ is $3dB$, the system
bandwidth $1$KHz and transmitting power is $15 dBW$. Suppose that
the distance between the transmitter and the receiver is $1000m$ and the pathloss exponent is $4$. In
particular, another very important parameter which will greatly
affect the simulation is the sampling interval, i.e., $\Delta \tau$.
As shown previously, \textit{Theorem} 1 is assured only if
$\Delta\tau=t/N\rightarrow0$. Therefore, the sampling interval
should be as small as possible, or say, for certain fixed
observation duration $t$, the number of samples, i.e., $N$, should
be as large as possible. We select the sampling interval as $0.1\mu
s$ namely, $N\ge10^7$ samples in one second, which is in good
agreement with the parameter in practical communication systems.

\begin{figure}[!t]
\centering
\includegraphics[width=3.0in]{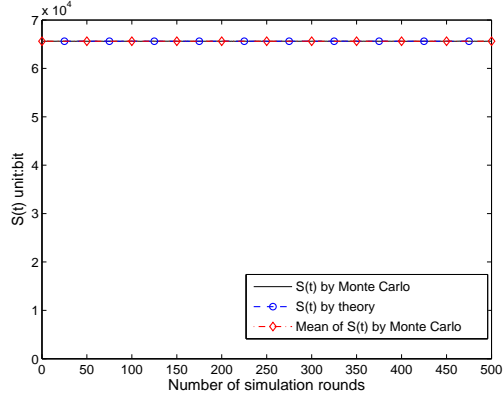}
\caption{The Rayleigh channel service process S(t) v.s. number of simulation
round. (Observation duration $t=5$s for each round.)}
\label{fig:horizontal_line}
\end{figure}

\begin{figure}[!t]
\centering
\includegraphics[width=3.0in]{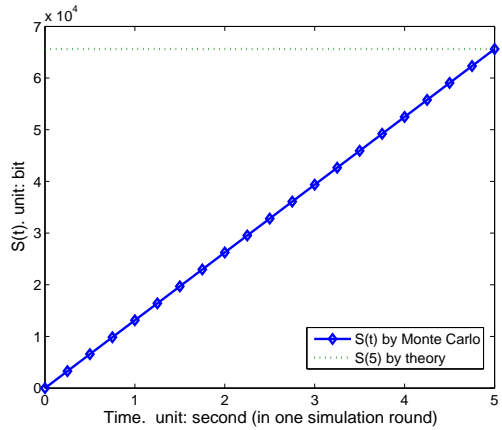}
\caption{The Rayleigh channel service process S(t) v.s. observation time}
\label{fig:slope}
\end{figure}

We consider the \textit{i.i.d.} Rayleigh fading channel and measure
the amount of service $S(t)$ by statistics with an observation
duration of $t=5s$ and run it independently for $500$ rounds. It is
observed in Fig.~\ref{fig:horizontal_line} that for certain fixed
moment such as $t=5s$, the variance of $S(t)$ is almost zero which
means that it is deterministic linear function of $t$. The
simulation result also fits the $c^\ast_{ra}$ by
(\ref{eq:st_rayleigh}) perfectly. {It can be further observed in
the detail when the vertical coordinate axis is zoomed in
that there are still small fluctuations. However,} the maximum deviation ratio of $S(t)$ from the average value in all of the simulation rounds is
less than $2.6\times10^{-5}$ {and will decrease when smaller $\Delta\tau$ is used}. Fig.\ref{fig:slope} illustrates the
channel service process $S(t)$ v.s. observation time $t$. It
confirms the deterministic time-linearity of the channel service
process $S(t)$ for each time $t$, which is consistent with our
theoretical analysis, namely, the amount of channel service
increases linearly with time $t$.

\begin{figure}[!t]
\centering
\includegraphics[width=3.0in]{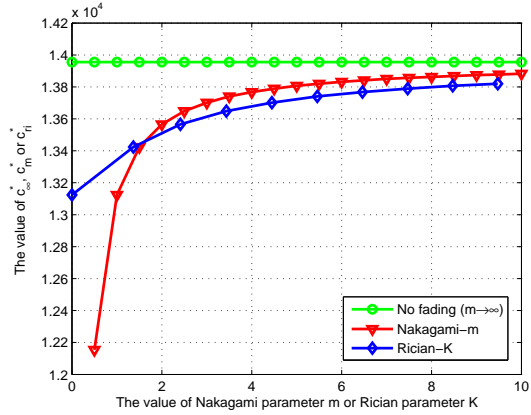}
\caption{The $c_m^\ast,c^\ast_{ri}$ v.s. m(\textrm{or} K)}
\label{fig:c_m_K}
\end{figure}

In Fig.\ref{fig:c_m_K}, we provide the $c_m^\ast$ v.s. Nakagami parameter $m$ and $c^\ast_{ri}$ v.s. Rician parameter $K$. As is known, the Nakagami channel with $m=1$ and the Rician channel with $K=0$ both reduce to the Rayleigh channel. It is observed in Fig. \ref{fig:c_m_K} that the $c^\ast_m|_{m=1}=c^\ast_{ri}|_{K=0}$ and they fit the service provided by a Rayleigh channel in 5 seconds, i.e., $S(t)|_{t=5}$, which is shown in Fig. \ref{fig:horizontal_line}. The AWGN channel capacity $c_{awgn}$ with a average LOS received power $Pr$ is also presented in Fig. \ref{fig:c_m_K}, and it can be seen that as $m$ (or $K$) increases, the $c^\ast_m$ and $c^\ast_{ri}$ increase accordingly and will convergence to $c^\ast_\infty =c_{awgn}$, which confirms {\em Corollary} 2 and {\em Corollary} 3.

By these simulation results, the deterministic time-linearity of the extensively investigated {\em i.i.d.} fading channels is verified once again. In the paper, $S(t)$ is defined as a stochastic integration and is investigated in the sense of mean square. Therefore, at any
time $t$, $S(t)$ is a constant other than a random variable, so is $\frac{S(t)}{t}$. To make the problem more trackable, we adopt in the paper a
sampling method to approximate it, i.e.,
$S(t)=\lim_{N\rightarrow\infty}\sum_1^N C(t_n)\Delta t$. During a
sampling time, it is assumed that the channel gain remains
unchanged, which is a commonly used processing method for
integration. When the sampling interval goes to zero, one can get
the integration value. More importantly, such a treating  will not change the inherent characteristic of $S(t)$. Besides, we adopt the blocking fading concept for the simplicity of expression. This is similar to the approximation of the Brown motion by random walking, where both the two items are stochastic processes. However, it is really a lucky thing and it can be proved in this paper that the variance of the independent increment process $S(t)$ is zero, i.e., $D[S(t)]=0$. This is totally different from $D[E[S(t)]]=0$ or $D[\frac{S(t)}{t}]=0$ and is a new result. As mentioned previously, there still are some points to be noted. The channel service considered is a concept of
large  time scales other than the fading property in small scales.
In fact, it holds when the ratio of the observation time and the
sampling time ($\frac{t}{\Delta t}$) is sufficiently large. And even for
a block fading channel, the result holds if the observation time $t$
is sufficiently large. However, if one investigates the channel service on smaller time scales, some physical layer technologies should be used to adapt to the instantaneous channel fluctuations, namely the fading characteristic
of the channel. To achieve this goal, buffers must be used at the
transmitter. In this way, data can be stored in the buffer at the
transmitter when the channel is in bad condition. For this topic, we
have obtained some results on the channel utilization and buffer
overflow probability, which will be presented in our following works.

\section{Conclusions}\label{sec:conclusions}

This work introduced a new picture of \textit{i.i.d.} fading
channels from the viewpoint of the cross layer. That is, we proved
that the channel service process $S(t)$ of an \textit{i.i.d.} fading
channel is a deterministic linear function of time $t$, by using the
CF approach based on the infinitely divisible law. This work
provides some significant insights in both theory and applications.
First, different from conventional ergodic capacity or outage
capacity, it asserts that the \textit{i.i.d.} fading channel has a
deterministic transmission ability. In other words, there exists a
coding scheme such that the receiver can decode with zero error
probability, if the transmitter maintains a constant data rate no
more than $c_p^\ast$ in the point of view from application layer.
Second, in opposite to conventional opinions, this work asserts that
the high layer queuing delay is assured to be zero as long as the
transmission data rate is less than $c_p^\ast$. Otherwise, the
queuing delay will be upper-bounded, which is determined by the
difference between the transmission data rate and $c_p^\ast$.

\section*{Acknowledgement}
Prof. P. Fan's work was partly supported by the the China Major
State Basic Research Development Program (973 Program) No.
2012CB316100(2) and National Natural Science Foundation of China
(NSFC) No. 61171064. Prof. K. B. Letaief's work was partly supported
by RGC under grant No. 610311.


\begin{thebibliography}{11}
\bibitem{MN-nakagami}
M. Nakagami, ``The m-Distribution ¨C A General Formula of Intensity Distribution of Rapid Fading,''
Statistical Methods in Radio Wave Propagation, Pergamon Press: Oxford, U.K., 1960, pp. 3-36.
\bibitem{MS-CWB}
McEliece R., Stark W., ``channels with block interference," IEEE
Trans. Inform. Theory, 1984 , vol. 30, no. 1,  pp. 44-53.
\bibitem{LA-Outage}
L. Li and A. J. Goldsmith, ``Capacity and optimal resource allocation
for fading broadcast channels: Part II: Outage capacity," IEEE
Trans. Inf. Theory, vol. 47, no. 3, pp. 1103-1127, Mar. 2001.
\bibitem{GS-Throughput}
G. Caire and S. Shamai(Shitz), ``On the achievable throughput of a
multiple- antenna Gaussian broadcast channel," IEEE Trans. Inf.
Theory, vol. 49, no. 7, pp. 1691-1706, Jul. 2003.
\bibitem{GP-CFC}
A. J. Goldsmith, P. P. Varaiya, ``Capacity of fading channels with
channel side information,¡± IEEE Trans. Inf. Theory, 1997, vol. 43,
no. 6, pp. 1986-1992.
\bibitem{BPS-FCI}
Biglieri E., Proakis J., Shamai, S., ``Fading channels:
information-theoretic and communications aspects," IEEE Trans.
Inform. Theory, 1998 , vol. 44, no. 6,  pp. 2619-2692.
\bibitem{DN-EC}
Dapeng Wu, Negi R., ``Effective capacity: a wireless link model for
support of quality of service," Wireless Communications, IEEE
Transactions on, 2003, vol. 2, no. 4, pp. 630-643.
\bibitem{QDP-EC-R}
Qing Wang, Dapeng Wu, Pingyi Fan, ``Effective capacity for a
Correlated Rayleigh Fading Channel," online published by Wiley
Wireless Communications and Mobile Computing.
\bibitem{QDP-EC-N}
Qing Wang, Dapeng Wu, Pingyi Fan, ``Effective capacity for a
Correlated Nakagami-m Fading Channel," online published by Wiley
Wireless Communications and Mobile Computing.
\bibitem{Levy}
James W. Pitman, ``Lecture notes probabily theory-Levy Process and
Infinitely Divisible Law," online available at
http://www.stat.berkeley.edu/users/pitman/s205s03/levy.pdf
\bibitem{GK-Notes}
Abbas El Gamal, Young-Han Kim, ``Lecture notes on network information
theory," Tsinghus University, spring, 2010.
\bibitem{Id}
``Infinite divisibility," online available at
http://en.wikipedia.org /wiki /Infinite $_-$ divisibility$_-$(probability)
\bibitem{G-WC}
Wireless Communications, Andrea Goldsmith, Stanford University,
pp.71-72, Cambridge University Press.
\end{thebibliography}
\end{document}